\documentclass[%
 eprint,
 superscriptaddress,
 amsmath,amssymb,
 aps,
 prb,
 longbibliography,
 lengthcheck,%
]{revtex4}

\usepackage{graphicx}
\usepackage{dcolumn}
\usepackage{bm}
\usepackage{color}

\newcommand\T{\rule{0pt}{2.6ex}}

\begin{document}

\title{Li intercalation in graphite: a van der Waals density functional study}
\author{E. Hazrati}
\email{e.hazrati@science.ru.nl}
\affiliation{Radboud University Nijmegen, Institute for Molecules and Materials, Heyendaalseweg 135, 6525 AJ Nijmegen, The Netherlands}
\author{G. A. de Wijs}
\email{g.dewijs@science.ru.nl}
\affiliation{Radboud University Nijmegen, Institute for Molecules and Materials, Heyendaalseweg 135, 6525 AJ Nijmegen, The Netherlands}
\author{G. Brocks}
\email{g.h.l.a.brocks@utwente.nl}
\affiliation{Computational Materials Science, Faculty of Science and Technology and MESA+ Institute for Nanotechnology, University of
             Twente, P.O. Box 217, 7500 AE, Enschede, The Netherlands}
\date{\today}

\begin{abstract}

Modeling layered intercalation compounds from first principles poses a problem, as many of their properties are determined by a subtle balance 
between van der Waals interactions and chemical or Madelung terms, and a good description of van der Waals interactions is often lacking. 
Using van der Waals density functionals we study the structures, phonons and energetics of the archetype layered intercalation compound Li-graphite. 
Intercalation of Li in graphite leads to stable systems with calculated intercalation energies of $-0.2$ to $-0.3$~eV/Li atom, (referred to 
bulk graphite and Li metal). The fully loaded stage 1 and stage 2 compounds LiC$_6$ and Li$_{1/2}$C$_6$ are 
stable, corresponding  to two-dimensional $\sqrt3\times\sqrt3$ lattices of Li atoms intercalated between two graphene planes. 
Stage $N>2$ structures are unstable compared to dilute stage 2 compounds with the same concentration. At elevated 
temperatures dilute stage 2 compounds easily become disordered, but the structure of
Li$_{3/16}$C$_6$ is relatively stable, corresponding to a $\sqrt7\times\sqrt7$ in-plane packing of Li atoms. First-principles calculations, along with a Bethe-Peierls model of finite temperature effects, allow for a microscopic description of the observed voltage profiles. 

\end{abstract}

\maketitle

\section{Introduction}
\label{Introduction}

Intercalation of metal atoms into graphite\cite{Dresselhaus.Dresselhaus} has lead to a wealth of interesting physical phenomena. Alkali, alkaline earth or rare earth metal atoms can be inserted between the 
graphene layers of graphite without disrupting the bonding pattern within the graphene layers, and the electronic structure of the 
metal-graphite compound can be deduced from the interactions between the graphene and the metal layers.\cite{PRL.106.187002,NatPhys.1.42} Intercalation in few-layer graphene is explored for modifying its electronic, transport, and optical properties.\cite{NatComm.4.2772,NatComm.5.4224} Some of these metal-graphite 
compounds even become superconducting.\cite{NatPhys.1.39,PRL.106.187002,PRB.87.075108}  The metal intercalation process is usually reversible, making graphitic carbon one of the most used materials in anodes of rechargeable batteries.\cite{Science.270.590,Adv.Mater.10.725,Carbon.38.183,Adv.Mater.21.2664}  

Li-graphite is the archetypical intercalation compound in this class, whose composition Li$_x$C$_6$ can easily be varied between $x=0$ and $x=1$, giving 
rise to a surprisingly rich phase diagram.\cite{PhysRevB.44.9170,JES140.2490,JES160.A3198,J.Phys.Chem.Solids.57-775,JES151.A422,Carbon.61.140} Li intercalation in carbon-based materials is also relevant to hydrogen storage, as a tool to manipulate dehydrogenation reactions.\cite{Gross2008a,Ngene2010cc,Liu2011c,JPCC.118.5102} The structure of the fully loaded stage~1 compound LiC$_6$ consists of graphene alternating with a layer of Li atoms. Controlling the Li content electrochemically and monitoring the Li$_x$C$_6$ potential as a function of $x$ shows a sequence of plateaus that is interpreted as subsequent phase 
equilibria.\cite{PhysRevB.44.9170,JES140.2490,JES160.A3198,J.Phys.Chem.Solids.57-775,JES151.A422,Carbon.61.140} Stage~$N$ ($=2,3,...$) defines a structure consisting of a stack of $N$ graphene layers alternating with a Li layer,\cite{Dresselhaus.Dresselhaus} and for $1/N<x<1/(N-1)$ it is proposed that the stage~$N$ and stage~$N-1$ phases are in equilibrium. Upon decreasing the Li content to $x<1/N$, one then moves to the next equilibrium plateau between stage~$N+1$ and stage~$N$ phases. This simple model is under scrutiny though,  as neutron 
diffraction experiments give evidence for the formation of phases with partially filled 
Li layers instead of fully completed higher order stage~$N$ phases,\cite{JES160.A3198} and calculations suggest the 
relative stability of certain partially filled structures.\cite{JPCC112.3982}

Experimental characterization of Li intercalation is hampered by kinetic barriers,\cite{PhysRevB.88.094304} which can give rise to non-equilibrium intermediate phases. First-principles calculations provide a valuable contribution to modeling the intercalation process,\cite{PhysRevB.83.144302,PhysRevB.86.035147} but specifically for the prototype intercalation compound Li-graphite this has proven to be a challenging task. Different Li-graphite phases emerge from a subtle balance between the interactions of the Li atoms with the graphene sheets and the van der Waals (vdW) interactions between the graphene sheets.\cite{PhysRevB.82.125416} The most widely used first-principles approaches, i.e. 
local\cite{PhysRev.140.A1133} or semi-local approximations\cite{PhysRevLett.77.3865} to density functional theory (DFT),\cite{PhysRev.136.B864} fail to describe the inherently non-local vdW interactions. 

The phase diagram of Li-graphite based upon calculations with a semi-local functional, without correcting for vdW interactions,
is even qualitatively wrong, as it does not yield any particularly stable ordered structure 
besides the stage~1 compound LiC$_6$, which is in contradiction to experimental 
results.\cite{PhysRevB.44.9170,JES140.2490,JES160.A3198,J.Phys.Chem.Solids.57-775,JES151.A422,Carbon.61.140}  Although it 
does not include vdW interactions, the local density approximation (LDA) yields reasonable 
equilibrium structures, both for graphite, as well as for the stage 1 intercalation 
compound LiC$_6$.\cite{PRB.68.205111,JPCM.25.445010} As we will discuss below, the energetics of intercalation 
is not described very accurately by LDA however. The interlayer binding energy of graphite is a factor of two too small, whereas 
the Li intercalation energy is a factor of two too large.

One may include vdW interactions by adding a parametrized semi-empirical atom-atom dispersion energy to the conventional Kohn-Sham 
DFT energy, as in the DFT-D2 method.\cite{JComChem.27.1787} A problem with this approach is that vdW interactions depend critically 
on the charge state of the atoms involved. For instance, the vdW interaction of a Li$^+$ ion with its environment is substantially 
smaller than that of the neutral Li atom (because virtual excitations from the 2s shell give a large contribution to the polarizability 
of the atom and the vdW interaction). As Li atoms interacting with graphene become partially ionized,\cite{JPCC-113-8997} the 
parameters describing the vdW interaction need to be refitted.\cite{Nano.Letters.12.4624} This means that the method loses its 
predictive power if the charge on the metal atoms is not known beforehand. Other semi-empirical schemes have been developed that are specifically targeted at modeling vdW interactions in layered materials such as graphite, requiring the input of the material's elastic properties, obtained either from advanced many-body calculations, or from experiment.\cite{JPCM.25.445010}
 
Many-body approaches such as quantum Monte Carlo (QMC) or the random phase approximation (ACFDT-RPA) incorporate a description 
of the vdW interactions, and have been used to calculate the binding between the graphene layers in graphite, for 
instance.\cite{PRL.103.196401,PRL.105.196401} However, as such methods are computationally very demanding, they cannot be applied to Li-graphite compositions that require the use of large unit cells. An alternative approach to include vdW interactions 
is using a van der Waals density 
functional (vdW-DF),\cite{PhysRevB.62.6997,PhysRevLett.91.126402,PhysRevLett.92.246401,PhysRevB.76.125112} which is an explicit 
non-local functional of the density. This is the approach we use here.

In this paper we study the intercalation of Li into graphite entirely from first-principles using a vdW DFT functional, i.e., without 
any empirical data or {\em ad hoc} vdW corrections.
First we validate this approach by calculations on pure graphite. In particular we show that the phonon band structure and elastic 
constants of graphite are reproduced well, including the ones that depend on the coupling between the graphene layers, where the 
contribution of vdW interactions is critical.
Then we apply this approach to intercalation compounds Li$_x$C$_6$, $0\leq x\leq 1$, identifying stable phases and their properties. 
We establish that the fully loaded stage 1 and stage 2 compounds are stable, but stage $N>2$ structures are unstable compared to 
dilute stage 2 compounds with the same concentration. At elevated temperatures these dilute stage 2 compounds easily become disordered.

This paper is organized as follows.
Sec.~\ref{computational} discusses the computational details.
In Sec.~\ref{bulk-graphite} we apply the vdW-DF approach to bulk graphite and compare the performance of
different versions of the vdW-DF.
In Sec.~\ref{Li-GS} we study the Li intercalation into graphite, and Sec.~\ref{discussion_and_conclusions} presents the summary and conclusions.

\section{Computational Methods}
\label{computational}

We perform first-principles calculations within the framework of density functional theory (DFT)\cite{PhysRev.136.B864,PhysRev.140.A1133} using
the projector augmented wave method (PAW)\cite{PhysRevB.50.17953,PRB59-1758} 
as implemented in the Vienna {\sl ab initio} simulation package (VASP).\cite{PRB54-11169,CMS6-15}
To include the non-local vdW interactions, we use a van der Waals density functional  
\cite{PhysRevLett.92.246401,PhysRevB.76.125112} as implemented in VASP\cite{J.Phys.Condens.Matter.22.022201,PhysRevB.83.195131} 
using the algorithm of Ref.~\onlinecite{PhysRevLett.103.096102}. The exchange-correlation energy in the vdW-DF has the form
\begin{equation}
E_{\rm xc} = E_{\rm x} + [E_{\rm c}{\rm (vdW)} + E_{\rm c}{\rm (loc)}],
\label{E-XC}
\end{equation}
where $E_{\rm c}{\rm (vdW)}$ is the energy resulting from non-local electron-electron correlations, approximated by an expression in 
terms of the electron density, \cite{PhysRevLett.92.246401,PhysRevB.76.125112} and $E_{\rm c}{\rm (loc)}$ represents the energy 
contribution of the local electron-electron correlations, for which the local density approximation (LDA) is used. In the 
original vdW-DF,\cite{PhysRevLett.92.246401} the revPBE functional\cite{PhysRevLett.80.890} is used to calculate the contribution 
of the exchange energy $E_{\rm x}$. We also try the vdW-DF2 functional,\cite{PhysRevB.82.081101} which  uses a modified vdW kernel 
along with the PW86 exchange functional.\cite{PhysRevB.33.8800} Both the original vdW-DF and the vdW-DF2 functionals tend to 
overestimate the lattice constants and underestimate the formation energies of solids somewhat.\cite{PhysRevB.83.195131} The 
optimized exchange functionals introduced in Refs.~\onlinecite{J.Phys.Condens.Matter.22.022201}
and \onlinecite{PhysRevB.83.195131}, i.e.\ optB88, optPBE and optB86b, alleviate these problems, and we will test these functionals. 
       
Standard PAW data sets are used, which are generated and unscreened using the PBE functional.\cite{PhysRevLett.77.3865}
For lithium we use an all-electron PAW description, whereas for carbon the 1s core state is kept frozen.
A kinetic energy cutoff of 550~eV is employed for the plane wave expansion of the Kohn-Sham states.
The atomic positions are optimized with the conjugate gradient method until the forces on atoms are less than 
10$^{-2}$~eV/\AA. This criterion is sufficiently strict to obtain converged total energies.
In addition to atomic positions, the volume and shape of the cells are optimized for bulk graphite and the 
Li-graphite compounds.\cite{note.volume}

Lattice vibrational frequencies are calculated for bulk graphite and the Li-graphite systems from the dynamical matrix, where 
the force constants are obtained using the finite difference 
method of Ref.~\onlinecite{EPLL}. Calculating an accurate dynamical matrix requires starting from very accurate atomic equilibrium 
positions. So as a first step the latter are further optimized until the forces on the atoms are less than 10$^{-4}$~eV/\AA. 
Next the atoms are displaced one-by-one and the resulting forces on all the other atoms are calculated.  The typical size of a 
displacement is $n \times 0.015$~\AA. Four displacements ($n=\{-2,-1,1,2\}$) per independent  degree-of-freedom are applied in 
order to remove anharmonic contributions to the forces.

A $\Gamma$-centered 24$\times$24$\times$10 $k$-point mesh is used to sample the Brillouin zone (BZ) of AB stacked graphite. 
The same $k$-point density is used for the calculations on Li intercalation in graphite. The Methfessel-Paxton (MP) 
scheme\cite{PhysRevB.40.3616} with a smearing width of 0.2~eV is employed for the occupation of the electronic levels. 
The energy convergence with respect to the $k$-point sampling is better than 1~meV/C.  

\section{Results}
\label{resultsection}

\begingroup
\squeezetable
\begin{table*}[htb!]
\caption{The equilibrium in-plane lattice constant $a$, interlayer distance $d$ and interlayer binding energy $E_{\rm B}$
         of graphite calculated using different exchange and correlation functionals, 
         compared to experiment (Exp.) and to results from many body wave function calculations (ACFDT-RPA, QMC).}
\begin{ruledtabular}
\begin{tabular}{ccccccccccc}
Exchange           &   PBE   &   PBE   &   optB88  &   optPBE  &   optB86b &   revPBE    &  rPW86       & ACFDT-RPA\footnotemark[1] & QMC\footnotemark[2] &  Exp.     \\
Correlation        &   PBE   & vdW+LDA &  vdW+LDA  &   vdW+LDA &   vdW+LDA &   vdW+LDA   &  vdW2+LDA    &                           &                     &           \\ 
\hline                                                                                                                          
$a$(\AA) \T        &   2.47  &   2.47  &   2.47    &   2.48    &   2.47    &   2.48      &   2.48        &           &                    &  2.46\footnotemark[3]     \\
$d$(\AA)           &   4.40  &   3.44  &   3.36    &   3.44    &   3.31    &   3.59      &   3.51        & 3.34      & 3.43               &  3.34\footnotemark[3]     \\
$E_{\rm B}$(meV/C) &   1.0   &  70.8   &  69.5     &  63.7     &  69.9     &  52.7       &  52.0         & 48        & 56$\pm$5           & 52$\pm$5\footnotemark[4]  \\       
\end{tabular}
\end{ruledtabular}
\footnotetext[1]{Ref.~\onlinecite{PRL.105.196401}}
\footnotetext[2]{Ref.~\onlinecite{PRL.103.196401}}
\footnotetext[3]{Ref.~\onlinecite{PhysRev.100.544}}
\footnotetext[4]{Ref.~\onlinecite{PhysRevB.69.155406}}
\label{graphite-properties}
\end{table*}
\endgroup

\subsection{Graphite}
\label{bulk-graphite}

We start with bulk graphite to critically test different vdW-DFs. Key quantities are the equilibrium structure and the 
equilibrium binding energy. Somewhat more demanding properties that probe the potential energy surface close the equilibrium 
minimum, are the phonon spectrum and the elastic constants. Table~\ref{graphite-properties} gives the equilibrium distance $d$ 
between the graphene layers and the equilibrium interlayer binding energy $E_{\rm B}$ (the graphite total energy subtracted 
from twice the total energy of isolated graphene layers), calculated using different exchange and vdW functionals. All tested 
functionals yield an in-plane lattice constant $a$ very close to the experimental value, indicating that the binding within 
a graphene plane is represented well. The interlayer distance $d$, however, is considerably overestimated by plain PBE without 
vdW forces (PBE-PBE): 4.40~\AA\ vs.\ 3.34~\AA. Indeed, the lack of vdW attraction is also apparent from a near absence of 
any interlayer binding ($E_{\rm B} = 1$~meV/C). LDA gives a reasonable interlayer distance of 3.25\AA, but a interlayer binding of only 24 meV.\cite{PRB.68.205111,JPCM.25.445010,PRL.103.196401,PRL.105.196401}

By including vdW interactions both the interlayer distance and binding energy are reproduced markedly better. There is a 
modest spread in the results produced by the different functionals. The optB88-vdW and optB86b-vdW functionals give the 
best performance regarding the structure, with optimized interlayer distances within 1\% of the experimental value (3.34~\AA).
The PBE-vdW and optPBE-vdW functionals give interlayer distances that are 3\% too large, and the interlayer distances produced 
by the revPBE-vdW and rPW86-vdW2 functionals are 5\% and 7.5\% too large, respectively. Concerning performance with regard to 
binding energy, the order of the functionals is reversed. The revPBE-vdW and rPW86-vdW2 functionals give a binding energy that 
is very close to the experimental value and to the value obtained from quantum Monte Carlo calculations (QMC).\cite{PRL.103.196401} 
The other functionals (PBE-vdW, optB88-vdW, optPBE-vdW and optB86b-vdW) overestimate the experimental binding energy by 21 
to 24~$\%$. These results are in line with previous findings.\cite{PRB.82.153412,PRB.85.205402,JPCM24.424216}

From here on we select the optB88-vdW functional for our calculations, as it gives a very good interlayer distance and an acceptable 
interlayer binding energy. Details of the graphite structure are also given correctly. For instance, AB-stacked graphite is 
10~meV/C more stable than AA-stacked graphite, which is in agreement with experiment.\cite{bernal_graphite,Europhys.Lett.28.403} 
Moreover, this result is in excellent agreement with the number of 10~meV/C obtained in a recent ACFDT-RPA calculation.\cite{PRL.105.196401}
Note that the optPBE-vdW functional, adopted in Ref.~\onlinecite{RSCAdv.4.4069.2014}, performs about equally well (7~meV/C).

\begin{figure}[t!]
\includegraphics{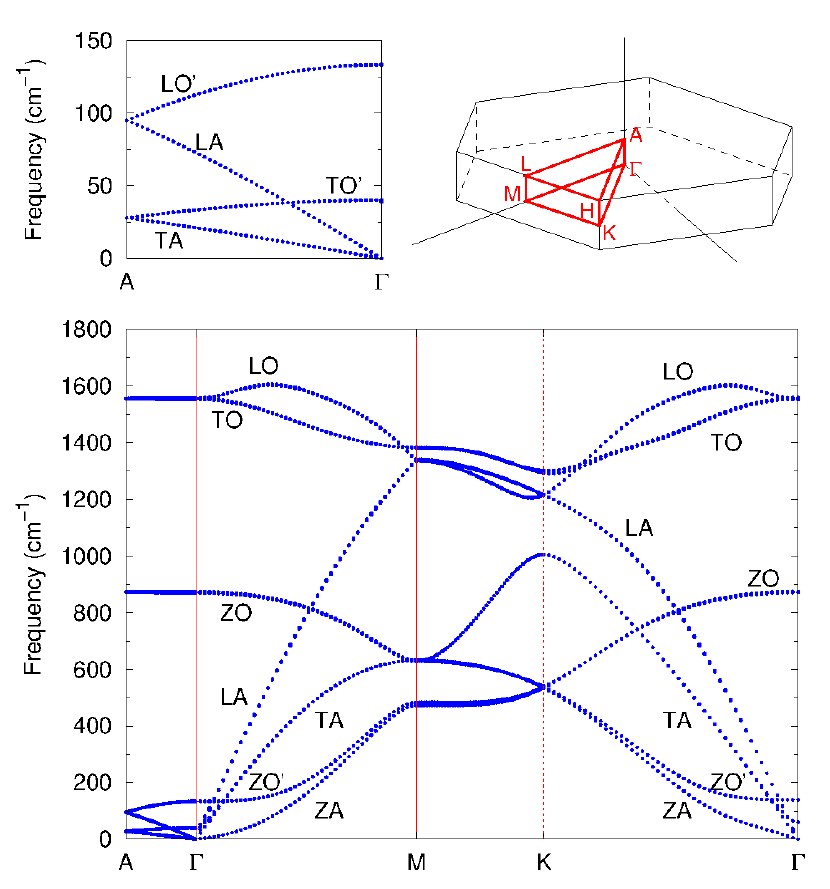}
\caption{(Color online) Graphite phonon dispersion calculated with the optB88-vdW functional starting from the optimized equilibrium structure.\cite{Note.cells}
On top left and right an enlargement of the low-frequency $\Gamma$-A region and the 
Brillouin zone with the high-symmetry points are shown, respectively.}
\label{fig-graphite}
\end{figure}

Phonons probe the potential energy surface close to the equilibrium structure, and are therefore a good test on the functional. Of particular interest are the low frequency phonons that involve interlayer motions, as vdW interactions 
play a major role there. Figure~\ref{fig-graphite} shows the graphite phonon dispersion calculated with the optB88-vdW 
functional, starting from the optimized equilibrium structure, i.e., the optimized in-plane lattice constant $a=2.47$\,\AA\ 
and interlayer distance $d= 3.36$\,\AA.   
The calculated phonon dispersions are in good agreement with 
experiments.\cite{PhysRevLett.92.075501,Oshima19881601,SIA:SIA1948,PhysRevB.55.7927,PhysRevB.5.4951}
This is evident from Table~\ref{graphite-phonon}, which lists phonon frequencies at the high-symmetry 
points $A$, $\Gamma$, $M$, and $K$, see also Ref.~\onlinecite{PhysRevB.71.205214}. The labels $L$, $T$ and $Z$ denote longitudinal, in-plane transversal and out-of-plane transversal 
polarization respectively. A primed O (O$'$) labels an optical mode where within the layers the atoms oscillate 
in phase whereas the two layers in the unit cell oscillate in anti-phase. An unprimed optical mode is a mode 
where atoms inside the same layer move in opposite directions.  

\begin{table}[tb!]
\caption{Phonon frequencies of graphite computed with the optB88 functional at the high-symmetry 
points $A$, $\Gamma$, $M$, and $K$ in cm$^{-1}$, compared to experimental results \cite{PhysRevLett.92.075501,Oshima19881601,SIA:SIA1948,
PhysRevB.55.7927,PhysRevB.5.4951,tuinstra:1126}.}
\begin{ruledtabular}
\begin{tabular}{lcc}
                         & optB88-vdW   &  Experiment      \\
\hline
$A\rm_{TA/TO'}$          &      28      &   35\footnotemark[1]            \\
$A\rm_{LA/LO'}$          &      95      &   89\footnotemark[1]            \\
$A\rm_{LO}$              &     873      &                 \\
$A\rm_{TO}$              &    1555      &                 \\
$\Gamma\rm_{LO'}$        &      40      &   49\footnotemark[1]            \\
$\Gamma\rm_{ZO'}$        &     139      &   95\footnotemark[2], 126\footnotemark[1]       \\
$\Gamma\rm_{ZO}$         &     870      &  861\footnotemark[2]            \\
$\Gamma\rm_{LO/TO}$      &    1553, 1558      & 1575\footnotemark[6], 1590\footnotemark[2]       \\
$M\rm_{ZA}$              &     471      &  471\footnotemark[1], 465\footnotemark[2], 451\footnotemark[4]  \\
$M\rm_{TA}$              &     628      &  630\footnotemark[4]            \\
$M\rm_{ZO}$              &     632      &  670\footnotemark[2]            \\
$M\rm_{LA}$              &    1335      & 1290\footnotemark[3]            \\
$M\rm_{LO}$              &    1340      & 1321\footnotemark[3]            \\
$M\rm_{TO}$              &    1383      & 1388\footnotemark[3], 1389\footnotemark[2]      \\
$K\rm_{ZA}$              &     534      &  482\footnotemark[4], 517\footnotemark[4], 530\footnotemark[5]  \\
$K\rm_{ZO}$              &     540      &  588\footnotemark[4], 627\footnotemark[5]       \\
$K\rm_{TA}$              &    1005      &                 \\
$K\rm_{LA/LO}$           &    1216      & 1184\footnotemark[3], 1202\footnotemark[3]      \\
$K\rm_{TO}$              &    1302      & 1313\footnotemark[4], 1291\footnotemark[5]      \\
\end{tabular}
\end{ruledtabular}
\footnotetext[1]{Ref.~\onlinecite{PhysRevB.5.4951}}
\footnotetext[2]{Ref.~\onlinecite{Oshima19881601}}
\footnotetext[3]{Ref.~\onlinecite{PhysRevLett.92.075501}}
\footnotetext[4]{Ref.~\onlinecite{SIA:SIA1948}}
\footnotetext[5]{Ref.~\onlinecite{PhysRevB.55.7927}}
\footnotetext[6]{Ref.~\onlinecite{tuinstra:1126}}
\label{graphite-phonon}
\end{table}
Note in particular that the low frequency modes (below $\sim 150$~cm$^{-1}$) between $\Gamma$ and $A$, which are 
particularly sensitive to the interlayer coupling, are well reproduced. This is not the case if one uses GGA (PBE-PBE) 
without vdW contributions, where the frequencies of the low-energy modes in particular are strongly 
underestimated.\cite{PhysRevB.71.205214} Forcing the experimental $c/a$ ratio upon the graphite structure largely repairs 
this deficit and yields sensible vibration frequencies.\cite{PhysRevB.71.205214} However, such a procedure requires input of experimental data.

Elastic properties are a second good test for the quality of the potential energy surface predicted by the first-principles 
calculations. Table~\ref{graphite-elastic} shows the elastic properties of graphite calculated with the optB88-vdW functional. 
To obtain the elastic constants we perform ground state total-energy calculations over a broad range of lattice 
parameters: $2.20\le a \le2.68$~\AA\ and $4.00\le c \le11.00$~\AA. 
The calculated results are then fitted to a two-dimensional sixth order polynomial.
The stiffness coefficients $C_{11}+C_{22}$, $C_{33}$ and $C_{13}$ are obtained as second derivatives of the energy 
with respect to $a$, $c$ and both $a$ and $c$, respectively.
The bulk modulus $B_{0}$ and the tetragonal shear modulus $C^{t}$ are obtained from the stiffness coefficients.
The procedure is similar to that of Ref.~\onlinecite{PhysRevB.71.205214}.

\begin{table}[tb!]
\caption{Elastic properties of graphite computed with the optB88 functional compared with the results from LDA, GGA, 
vdW-DF (revPBE) and RPA calculations, as well as with experiments. All data are in GPa}
\begin{ruledtabular}
\begin{tabular}{cccccc}
                         & $C_{11}+C_{22}$ & $C_{33}$ & $C_{13}$ & $B_{0}$ & $C^{t}$ \\
\hline
this work & 1200 & 35 & $-6.7$ & 33 & 216 \\
GGA\footnotemark[2] & 1230 & 45 & $-4.6$ & 41.2 & 223 \\
optB88-vdW\footnotemark[3] &  & 38 &  &  &   \\
revPBE-vdW\footnotemark[4] &  & 27 &  &  &  \\
ACFDT-RPA\footnotemark[5] &  & 36 &  &  &  \\
Exp. (300~K) & 1240$\pm$40\footnotemark[6] & 36.5$\pm$1\footnotemark[6] & 15$\pm$5\footnotemark[6] & 35.8\footnotemark[7] & 208.8\footnotemark[7]  \\
\end{tabular}
\end{ruledtabular}
\footnotetext[2]{Ref.~\onlinecite{PhysRevB.71.205214} with experimental $c$/$a$ ratio.}
\footnotetext[3]{Ref.~\onlinecite{JPCM24.424216}}
\footnotetext[4]{Ref.~\onlinecite{PhysRevB.76.155425}} 
\footnotetext[5]{Ref.~\onlinecite{PRL.105.196401}}
\footnotetext[6]{Ref.~\onlinecite{elastic}}
\footnotetext[7]{Ref.~\onlinecite{blakslee:3373}}
\label{graphite-elastic}
\end{table}

Table~\ref{graphite-elastic} compares our calculated elastic constants to experimental results,\cite{elastic,blakslee:3373} 
as well as to results obtained from GGA, vdW-DF (revPBE-vdW) and RPA calculations.\cite{PhysRevB.71.205214,PhysRevB.76.155425,PRL.105.196401}    
The elastic constant $C_{33} \propto \partial^2 E/ \partial c^2$ probes the interlayer interaction and is sensitive to the vdW interactions.
Our value is in very good agreement with experiment and with the result obtained from a ACFDT-RPA calculation.\cite{elastic,PRL.105.196401} 
It is a definite improvement over GGA [PBE-PBE] results (even when imposing the experimental $c/a$ ratio).\cite{PhysRevB.71.205214} 
A similar improvement is observed for the bulk modulus $B_0$. 

Note that the revPBE-vdW functional gives a $C_{33}$ that is somewhat too small, compared to experiment. Apparently, the revPBE-vdW 
functional gives an energy curve for the binding between the graphene layers that is somewhat too shallow, which is consistent with 
the fact that the revPBE-vdW equilibrium distance is somewhat too large, see Table~\ref{graphite-properties}. In this respect the 
optB88-vdW functional performs better, although it gives an interlayer binding energy that is somewhat too large. In fact, all 
elastic constants obtained with optB88-vdW are in good agreement with experiment, except for $C_{13}$. 

\subsection{Li intercalation}
\label{Li-GS}

The intercalation of Li in graphite leads to compounds Li$_x$C$_6$ with $0\leq x \leq 1$ with different structures as a function of the Li content $x$. We consider a large number of possible  Li$_n$C$_m$ ($x=6n/m$) structures and compositions, see Sec.~\ref{subsectionli2c6}, where 
we use the optB88-vdW functional in all calculations, unless explicitly mentioned otherwise. In all cases the cell parameters, as well 
as the atomic positions, are optimized. The intercalation energy $E_{\rm int}$ per Li atom is defined as 
\begin{equation}
E_{\rm int}({\rm Li}_n{\rm C}_m) = 
 \frac{1}{n}E({\rm Li}_n{\rm C}_m) - E({\rm Li_{\rm metal}}) - \frac{m}{4n}E_{\rm Gr} ,
\label{interc-energy}
\end{equation}
where $E({\rm Li}_n{\rm C}_m)$ is the total energy per formula unit of the Li$_n$C$_m$ (Li$_{6n/m}$C$_6$) phase, $E({\rm Li_{\rm metal}})$ 
is the total energy per atom of bcc bulk Li, and $E_{\rm Gr}$ is the total energy of one graphite unit cell (containing four carbon atoms). 
Alternatively the intercalation energy can be referred to the free Li atom by
subtracting the cohesive energy of the Li metal (1.578 eV/Li atom with optB88-vdW). Note that a negative value for $E_{\rm int}$ 
means that the intercalated compound is stable with respect to graphite and Li metal. 

\subsubsection{$\mathrm{LiC}_6$ and $\mathrm{Li}_{0.5}\mathrm{C}_6$}

We start with the fully loaded stage 1 compound LiC$_6$ and stage 2 compound Li$_{0.5}$C$_6$ (LiC$_{12}$). The stage 1 compound 
has -A-Li-A-Li- stacking with an optimized graphene interlayer distance of 3.64~\AA, which is close to the experimental value 
of 3.70~\AA.\cite{J.Electrochem.Soc.142.371} For the fully lithiated stage~2 compound  we consider 
both -A-Li-A-A-Li-A- and -A-Li-A-B-Li-B- stacking, and find that, in agreement with experiment,\cite{J.Phys.Chem.Solids.57-775} 
the former is favored over the latter. The calculated difference in intercalation energy is 32~meV/Li. The optimized average 
distance between the graphene layers in LiC$_{12}$  is 3.49~\AA, and the distance between the empty graphene layers is 3.27~\AA. 
These numbers are in good agreement with the experimental values of 3.51~\AA\ and 3.27~\AA, 
respectively.\cite{J.Phys.Chem.Solids.57-775,J.Electrochem.Soc.142.371} Evidently the optB88-vdW functional accurately reproduces 
the structures of LiC$_6$ and Li$_{0.5}$C$_6$.

\begin{figure}[tb!]
\includegraphics[width=8.5cm]{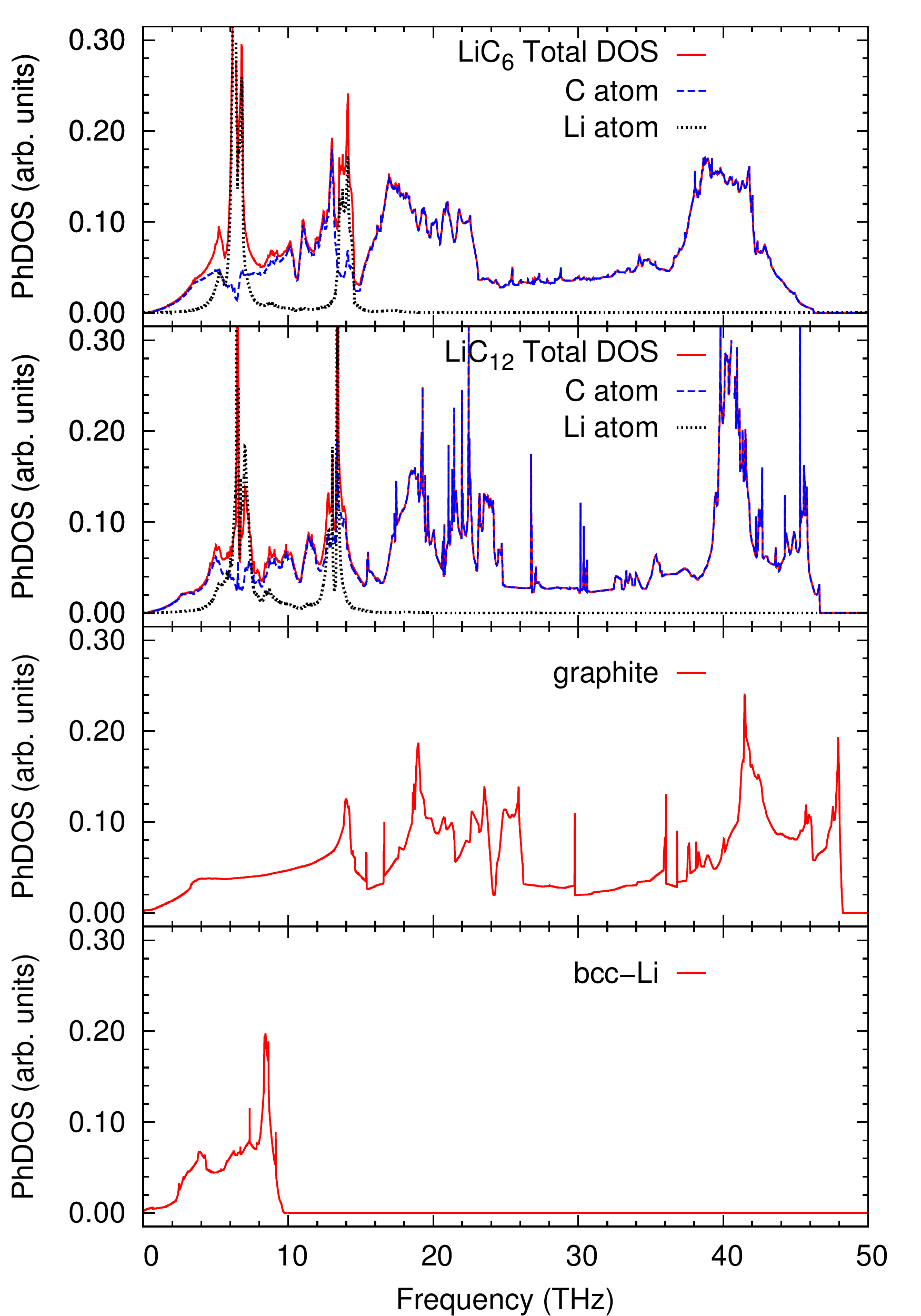}
\caption{(Color online) From top to bottom: phonon density of states (PhDOS) of stage 1 compound LiC$_6$, stage 2 compound LiC$_{12}$, pure 
graphite, and bcc Li metal.
The red line gives the total PhDOS, and the blue and black lines give the 
contributions of, respectively, the carbon and lithium atoms to the normal modes.}
\label{phononDOS}
\end{figure} 

The calculated intercalation energies for the stage 1 and stage 2 compounds LiC$_6$ and LiC$_{12}$ are $-0.217$ 
and $-0.273$~eV/Li, respectively, indicating the relative stability of the stage 2 compound. The intercalation free energies of 
the stage 1 and 2 compounds, extracted from electrochemical measurements at 300~K, are $-0.156$ 
and $-0.227$~eV/Li, respectively.\cite{JES140.2490} The intercalation enthalpies of LiC$_6$ and LiC$_{12}$ obtained from 
calorimetric measurements at 455~K with respect to liquid Li, are $-0.144$ and $-0.257$~eV/Li, respectively.\cite{J.Phys.Chem.Solids.57-947} 
Converting to solid Li as a reference state,\cite{fusionconvert,JAC-439-258} these enthalpies become $-0.113$ and $-0.226$~eV/Li. 
Even without including vibrational and finite temperature effects (to be discussed below), the calculations give intercalation energies that 
are consistently more negative than those obtained 
experimentally.\cite{note.LDA}
Part of this might be due to an error we make in 
describing the Li metal. For instance, the atomization energy of the Li metal comes out 0.1~eV too small with the 
optB88-vdW functional.\cite{PhysRevB.83.195131} 

So far we have not considered the vibrational contributions. The calculated phonon densities of states (PhDOS) of LiC$_6$ and 
LiC$_{12}$ are given in Fig.~\ref{phononDOS}. They can be compared to the PhDOSs of pure graphite and bulk Li metal. Whereas the 
phonon spectrum of graphite includes frequencies of up to 50 THz, see also Fig.~\ref{fig-graphite} and Table~\ref{graphite-phonon}, the 
phonon frequencies in bulk Li are all below 10 THz. The PhDOSs of LiC$_6$ and LiC$_{12}$ reflect this division into two 
frequency regimes. The low frequency modes definitely have a mixed carbon lithium character, whereas in the high frequency 
modes only carbon atoms participate. Comparing to the pure graphite and bulk Li spectra there are significant changes, however. 

\begin{figure}[tb!]
\includegraphics[width=8.5cm]{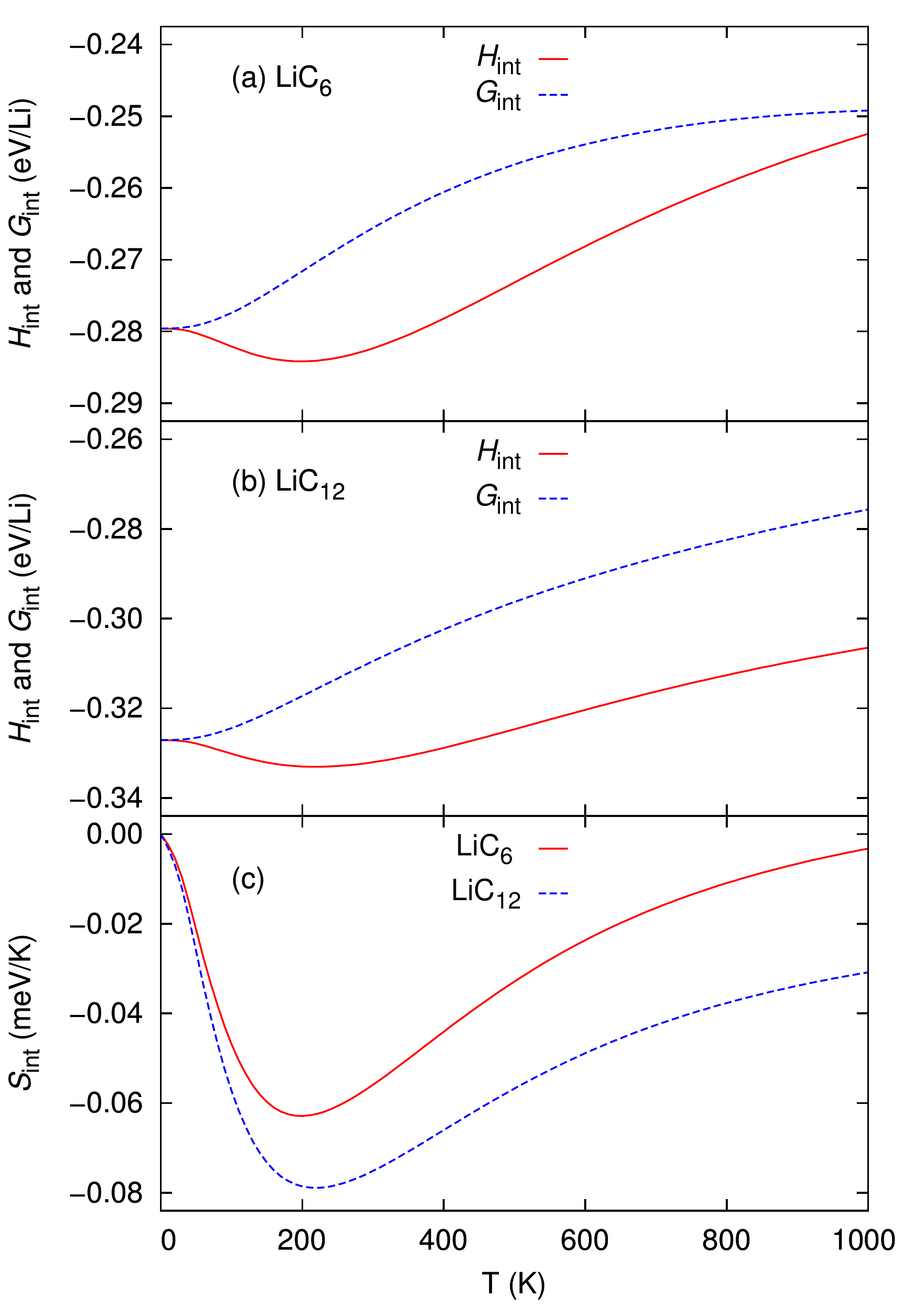}
\caption{(Color online) The intercalation enthalpy $H_\mathrm{int}$, see Eq.~(\ref{interc-energy}), entropy $S_\mathrm{int}$, and free 
energy $G_\mathrm{int}=H_\mathrm{int}-TS_\mathrm{int}$, of the stage 1 compound LiC$_6$ and the stage 2 compound 
LiC$_{12}$, including the phonon contributions.}
\label{enthalpy_entropy}
\end{figure}

The PhDOS at high frequencies of lithiated graphite is clearly shifted to lower frequencies, as compared to the PhDOS 
of pure graphite. Upon Li intercalation the in-plane C-C bond length becomes larger and the bonds become weaker, as Li 
atoms donate electrons to the $\pi^*$ anti-bonding states of graphite. This leads to lower vibrational C-C stretch 
frequencies, which is noticeable in the high frequency range. There are also changes in the low frequency range, where 
vibrational modes concerning the motion of Li atoms are found. A double peak structure in the PhDOS between 6 and 14 THz can 
be identified, and assigned to modes where the Li atoms vibrate in the $ab$-plane, or along the $c$-axis, with the latter 
vibrations having the highest frequency. On average, the vibrational frequencies of intercalated Li atoms clearly 
are larger than those in bulk Li, indicating that intercalation confines the motion of the Li atoms. 

Zero-point vibrational energies (ZPEs) are dominated by high frequency modes, which in this case are the stretch modes 
of the carbon lattice. As the frequencies of such modes are lower in intercalated graphite than in pure graphite, it 
means that the ZPE in intercalated graphite is lower. Hence the ZPE gives a negative contribution to the intercalation energy. 
Indeed, including zero-point vibrational energies (ZPEs) changes the intercalation energies by $-0.04$ and $-0.05$~eV/Li 
for LiC$_6$ and LiC$_{12}$, respectively. 

The temperature dependence of the intercalation free energy of LiC$_6$ and LiC$_{12}$ is also determined by the phonons, as 
there is no contribution from configurational entropy in these fully lithiated compounds. The vibrational energy and entropy 
contributions to the intercalation enthalpy, entropy, and free energy can be calculated using standard harmonic oscillator 
expressions.\cite{vanSetten2008,pccp.13.6043} The thermodynamic quantities are shown in Fig.~\ref{enthalpy_entropy}. The intercalation 
enthalpy hardly changes over the temperature range 0-400 K. This makes sense as the vibrational contributions are dominated by 
the high frequency modes, and the occupancy of these modes is not very sensitive to the temperature in this range. Note that the 
intercalation entropy is negative, and goes through a distinct minimum around 200 K. The entropy is dominated by the 
low-frequency modes. The stiffening of the Li vibrational modes in intercalated graphite (with respect to Li metal) reduces the 
entropy, resulting in a negative intercalation entropy. This effect has been observed 
experimentally.\cite{JES151.A422,J.pow.sources.119.850} 
Adding enthalpy and entropy contributions yields an intercalation free energy that is monotonically increasing with temperature.

The ZPE contribution to the intercalation energies of Li$_x$C$_6$ is almost constant for $x\gtrsim 0.375$, and one can 
expect it to be smaller for $x<0.375$. In the following we compare relative intercalation energies for different~$x$. The ZPE 
contribution is then relatively unimportant, hence we do not consider it from here on.

\subsubsection{$\mathrm{Li}_{x}\mathrm{C}_6$; $x < 0.5$}
\label{subsectionli2c6}

\begin{figure}[tb!]
\includegraphics[width=8.5cm]{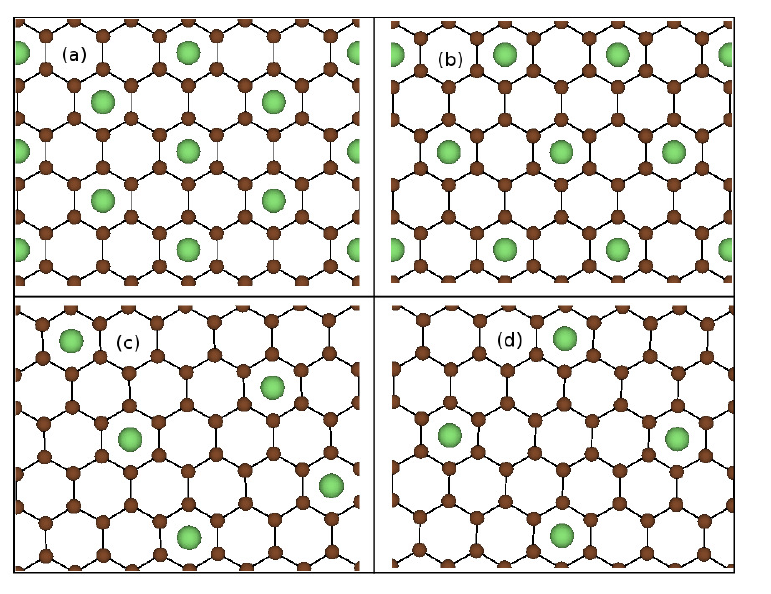}
\caption{(Color online) Examples of stage 2 compounds with different in-plane Li ordering: (a) LiC$_{12}$ 
with $\sqrt3\times \sqrt3$ Li ordering, (b) LiC$_{16}$ with $2\times 2$ Li ordering, (c) LiC$_{24}$ 
with $\sqrt7\times 2$ Li ordering, and (d) LiC$_{32}$ with $\sqrt7\times \sqrt7$ Li ordering. For 
simplicity, only AA stackings are shown.} 
\label{Li-struct}
\end{figure}

\begin{table}[tb!]
\caption{Optimized structures of selected stage 2 compounds LiC$_m$; $m$ $\ge$ 12 (Li$_x$C$_6$) with $n\times m$ in-plane unit cells, 
stage 1 compound LiC$_6$, and graphite C$_6$;
$d_{\rm av}$ is the average interlayer distance, in \AA;
$E_{\rm int}$ is the intercalation energy in eV/Li (without ZPEs).}
\begin{ruledtabular}
\begin{tabular}{ccccccccc}
             &     $x$    &    stack   &       $n\times m$      &        $d_{\rm av}$   &   $E_{\rm int}$    \\ \hline              
  C$_6$      &     0      &    AB      &       1$\times$1       &        3.36           &    0               \\
LiC$_{48}$   &     1/8    &    AABB    &  $\sqrt7\times4      $ &        3.48           &   $-0.256$         \\ 
LiC$_{40}$   &     3/20   &    AABB    &  $\sqrt7\times\sqrt{13}$ &      3.49           &   $-0.274$         \\ 
LiC$_{36}$   &     1/6    &    AABB    &  $\sqrt7\times3      $ &        3.50           &   $-0.275$         \\ 
LiC$_{32}$   &     3/16   &    AABB    &  $\sqrt7\times \sqrt7$ &        3.48           &   $-0.282$         \\ 
LiC$_{24}$   &     1/4    &    AABB    &  $\sqrt7\times 2     $ &        3.51           &   $-0.270$         \\ 
LiC$_{20}$   &     3/10   &    AABB    &  $\sqrt3\times \sqrt7$ &        3.51           &   $-0.263$         \\ 
LiC$_{20}$   &     3/10   &    AA      &  $\sqrt3\times \sqrt7$ &        3.54           &   $-0.239$         \\ 
LiC$_{16}$   &     3/8    &    AABB    &  $2     \times 2     $ &        3.53           &   $-0.263$         \\ 
LiC$_{16}$   &     3/8    &    AA      &  $2     \times 2     $ &        3.53           &   $-0.261$         \\ 
LiC$_{16}$   &     3/8    &    AA      &  $\sqrt3\times 2     $ &        3.53           &   $-0.257$         \\ 
LiC$_{12}$   &     1/2    &    AA      &  $\sqrt3\times \sqrt3$ &        3.49           &   $-0.273$         \\ 
LiC$_{6 }$   &     1      &    AA      &  $\sqrt3\times \sqrt3$ &        3.65           &   $-0.217$         \\ 
\end{tabular}
\end{ruledtabular}
\label{Li-GIC}
\end{table}

Whereas the structures of the fully lithiated stage 1 and stage 2 compounds are experimentally well established, less is known 
about the possible structures of Li$_x$C$_6$; $x < 0.5$. As a first step, we have constructed a number of dilute stage-2 
structures with compositions Li$_x$C$_6$, $1/8 \leq x \leq 1/2$, and $n \times m$ in-plane Li 
lattices, $\sqrt{3} \leq n,m \leq 4$. Examples of such structures are shown in Fig.~\ref{Li-struct}. The calculated optimized 
structural properties and intercalation energies of selected structures are listed in Table~\ref{Li-GIC}. As all these energies 
are negative, it follows that intercalation is favorable at any Li concentration. 

Li intercalation in graphite becomes more favorable upon increasing the concentration up to $x=3/16$. In the concentration 
range $3/16 < x\leq 1$ Li intercalation becomes slightly less favorable.
The average interlayer spacing $d_{\rm av}$ tends to increase with the Li concentration $x$. Exceptions are $x=3/16$ and $x=1/2$, which
coincide with minima in $E_{\rm int}(x)$. For these compositions that yield particular stable structures, $d_{\rm av}$ is smaller than that of adjacent compositions.
Note that the most stable stacking of the graphene planes is 
AA-type for $x > 3/8$, i.e., -A-Li-A-A-Li-A-. The stacking changes to AABB-type (-A-Li-A-B-Li-B-) for lower Li concentrations, 
however. Whereas the intercalation energies of AA and AABB stackings are within 2~meV/Li of one another for $x=3/8$, the 
difference increases to 150~meV/Li in favor of the AABB stacking for $x=1/8$.

\begin{figure}[tb!]
\includegraphics[width=6.0cm, angle = -90]{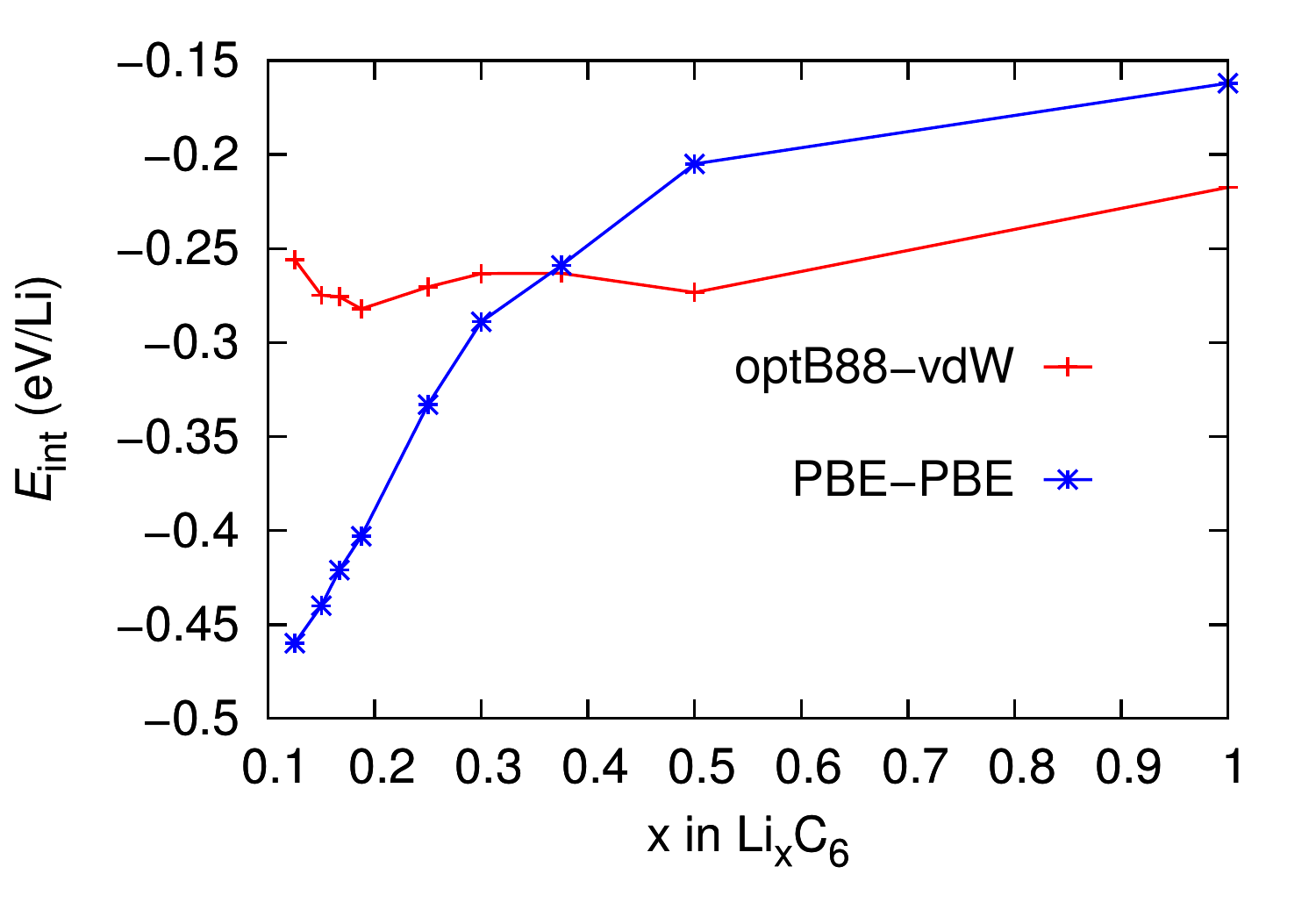}
\caption{(Color online) The intercalation energy $H_\mathrm{int}$, see Eq.~(\ref{interc-energy}) of the dilute stage 2 compounds 
Li$_x$C$_{6}$, calculated with the optB88-vdW (red) and PBE-PBE (blue) functionals.}
\label{optB88-PBE}
\end{figure}

To stress the importance of vdW interaction for the energetics of intercalation, Fig.~\ref{optB88-PBE} shows the 
Li intercalation energies calculated with the optB88-vdW and the PBE-PBE functionals (using optB88-vdW geometries).
As the latter functional 
lacks vdW interactions that give the interlayer bonding in graphite, intercalation of any amount of Li lowers the energy, as 
Li binds to the graphene planes. The bonding is partially ionic as Li donates electrons to the carbon lattice.\cite{JPCC-113-8997} 
The Coulomb repulsion between Li atoms/ions can be minimized in a diluted intercalation structure, which means that in absence 
of vdW interactions the intercalation energy monotonically increases with Li concentration. However, vdW interactions between 
the graphene planes oppose this trend. Intercalation disrupts the stacking of graphene planes, so vdW interactions prefer to 
cluster Li atoms such as to minimize the spatial extend of these disruptions.  

Omitting vdW interactions thus leads to an net overestimation of the effect of Li-graphene attractions in Li$_x$C$_6$ compounds 
with small $x$ and a net overestimation of the Li-Li repulsions for large $x$. Hence, the intercalation energy is too 
small (i.e., too negative) for small $x$, and too large for large $x$. The PBE-PBE intercalation energy is a monotonically increasing 
function of $x$, instead of having minima at a specific $x$. It means that PBE-PBE yields Li$_x$C$_6$ compounds where the Li concentration $x$ is a simple monotonic function of the Li chemical potential,  like in a simple lattice gas. This is clearly at variance with experiment, where phases with specific compositions are found to be thermodynamically stable.\cite{PhysRevB.44.9170,JES140.2490,JES160.A3198,J.Phys.Chem.Solids.57-775,JPCC112.3982} 

The balance between the graphene-graphene vdW bonding and the Li-graphene bonding gives the optB88-vdW curve 
shown in Fig.~\ref{optB88-PBE}. The curve has two shallow minima at concentrations $x=3/16$ and $x=1/2$, respectively. 
The latter corresponds to the fully loaded stage 2 compound, where the Li atoms order in plane in a regular $\sqrt3\times\sqrt3$ 
lattice, as shown in Fig.~\ref{Li-struct}. The $x=3/16$ structure corresponds to a dilute stage 2 compound, where the Li 
atoms order in plane in a regular $\sqrt7\times\sqrt7$ lattice, see Fig.~\ref{Li-struct}. One should note however that 
several other dilute stage 2 structures with compositions $x\leq 0.5$ have an intercalation energy within 20 meV 
of the two structures of Fig.~\ref{Li-struct}. We will come back to this point later.

Such dilute stage 2 structures, where partially loaded layers alternate with empty layers, are in fact more stable than stage 3-5 
structures of the same composition Li$_x$C$_6$, where fully loaded layers are separated by more than one empty layer. The 
calculated optimized structural properties and intercalation energies of selected stage 3-5 structures are listed in 
Table~\ref{Li-GIC2}. It means that, according to the calculations, it is not likely that stage 3-5 structures are 
formed during loading of graphite with Li. 

\begin{table}[tb!]
\caption{As Table~\ref{Li-GIC} but for stage N compounds LiC$_m$ with $\sqrt3\times \sqrt3$ in-plane unit cells.}
\begin{ruledtabular}
\begin{tabular}{ccccccccccc}
             &    $x$     &    stack   &       N                &    $d_{\rm av}$   &   $E_{\rm int}$  \\ \hline              
  C$_6$      &     0      &    AB      &       -                &    3.36           &    0             \\
LiC$_{30}$   &     1/5    &    AABAB   &       5                &    3.44           &  $-0.219$         \\
LiC$_{24}$   &     1/4    & AABABBAB   &       4                &    3.44           &   $-0.243$         \\
LiC$_{18}$   &     1/3    &    AAB     &       3                &    3.46           &   $-0.242$         \\
LiC$_{12}$   &     1/2    &    AA      &       2                &    3.49           &   $-0.273$         \\
LiC$_{6 }$   &     1      &    AA      &       1                &    3.65           &   $-0.217$         \\
\end{tabular}
\end{ruledtabular}
\label{Li-GIC2}
\end{table}

\subsubsection{Stable phases}
\label{subsectionstablephases}

\begin{figure*}[tb!]
\includegraphics[width=10.5cm,angle=-90]{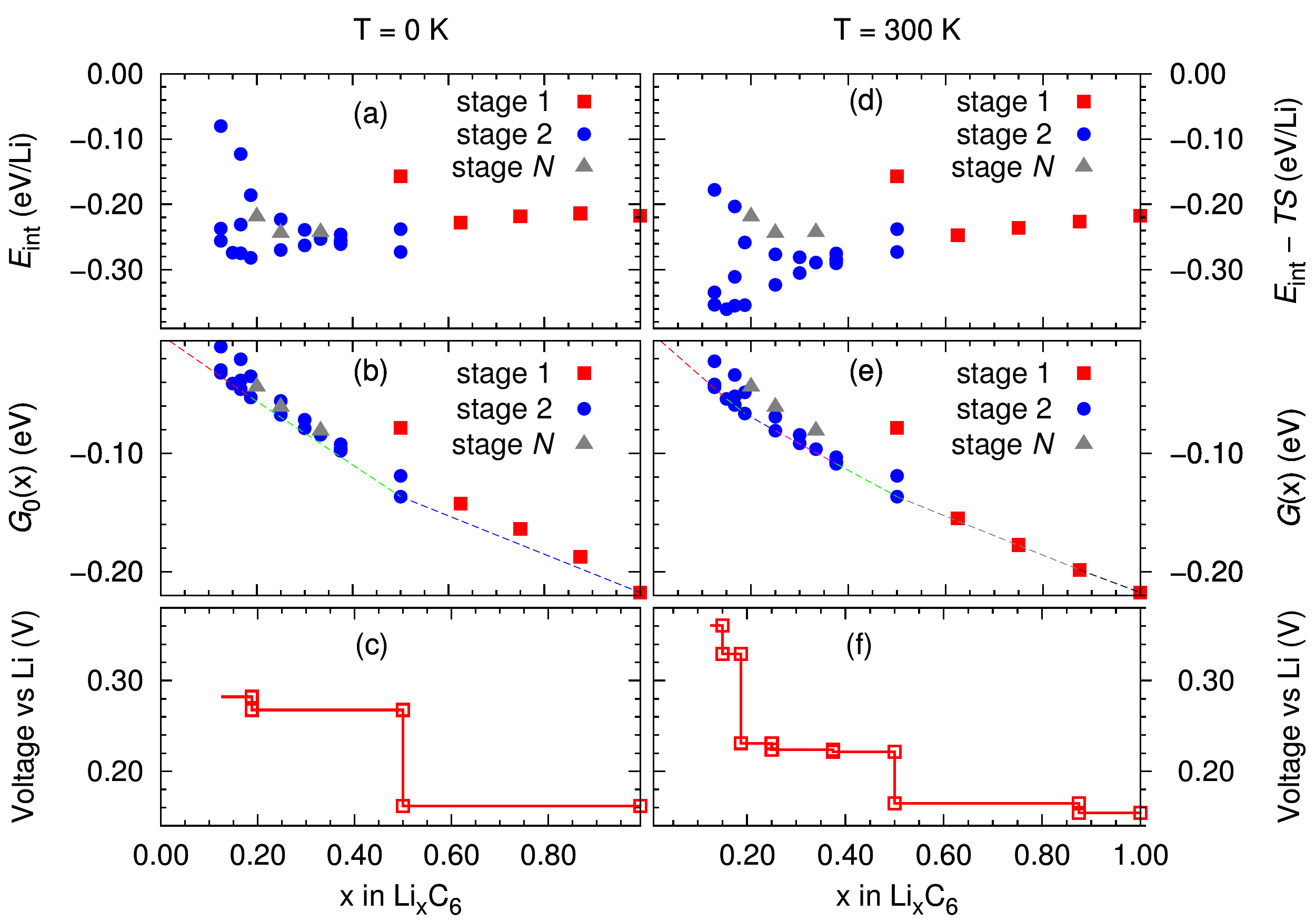}
\caption{(Color online) (a) Intercalation energy (eV/Li) versus concentration $x$ in Li$_x$C$_6$ structures (b) Zero temperature (free) 
energy $G_0(x)$ (eV) of Li$_x$C$_6$ structures versus Li concentration $x$; (c) Calculated zero temperature voltage 
profile of Li$_x$C$_6$ structures versus Li concentration $x$.
(d) Intercalation free energy (eV/Li) versus concentration $x$ in Li$_x$C$_6$ structures at room 
temperature ($T=300$ K) (e) Room temperature (free) energy $G(x)$ (eV) of Li$_x$C$_6$ structures versus Li 
concentration $x$; (f) Calculated room temperature voltage profile of Li$_x$C$_6$ structures versus Li concentration $x$.}
\label{Li-interc}
\end{figure*}

Intercalation energies for a large number of structures and different compositions are given in Fig.~\ref{Li-interc}(a). 
In agreement with the results shown in the previous subsection the two stage 2 Li$_x$C$_6$ structures with $x=3/16$ and $x=1/2$ give the optimal intercalation, corresponding to in-plane $\sqrt7\times\sqrt7$ and $\sqrt3\times\sqrt3$ orderings of Li 
atoms, respectively. Several dilute stage 2 structures with other compositions and slightly different in-plane orderings have 
slightly less favorable intercalation energies, but very different stage 2, or stage 1 and 3-5 structures have 
unfavorable intercalation energies.

A structure Li$_x$C$_6$ is stable with respect to decomposition into Li$_{x_1}$C$_6$ and Li$_{x_2}$C$_6$, $x_1 < x < x_2$, if 
its Gibbs free energy $G(x)$ is lower than that of the possible decomposition mixture, $[(x_2-x)G(x_1)+(x-x_1)G(x_2)]/(x_2-x_1)$. 
First we will consider zero temperature, where for solid states Gibbs free energies can be approximated by 
ground-state total energies,\cite{J.Pow.S.68.664}
\begin{equation}
G_0(x)\approx xE_\mathrm{int}(\mathrm{Li}_n\mathrm{C}_m),\; x=6n/m, \label{eq:Gx}
\end{equation}
where we use graphite and Li metal as reference phases. The values of $G_0$ are given in Fig.~\ref{Li-interc}(b) for the 
different structures and compositions~$x$. Constructing a convex curve from straight line segments between the data points with $x$ that 
are lowest in energy, only the points on that curve represent stable phases. The segments represent the free energies of 
decomposition mixtures. From our calculations the only stable phases at $T=0$ are then the stage~2 compounds LiC$_{32}$ and 
LiC$_{12}$, and the stage~1 compound LiC$_6$, corresponding to $x=3/16$, 1/2 and 1, respectively. 

Note that after starting the intercalation, first the phase appears with the lowest intercalation 
energy $E_\mathrm{int}$, i.e., LiC$_{32}$ ($x=3/16$). The following sequence of phases is then expected upon increasing 
the Li content. For $0 < x < 3/16$, graphite and the stage 2 compound LiC$_{32}$ coexist (red line in 
Fig.~\ref{Li-interc}(b)), followed by a coexistence of the stage 2 compounds LiC$_{32}$ and 
LiC$_{12}$ for $3/16 < x < 1/2$ (green line), and finally a coexistence of the stage 2 compound LiC$_{12}$ and the 
stage 1 compound LiC$_6$ for $1/2 < x <1$ (blue line). Experimentally the stability and structures of the LiC$_{12}$ 
and LiC$_6$ compounds are well established.\cite{PhysRevB.44.9170,JES140.2490,JES160.A3198,J.Phys.Chem.Solids.57-775,JPCC112.3982} 
Also quite consistently a stable phase with a composition around $x\approx0.2$ is observed, which we attribute to 
the $\sqrt7\times\sqrt7$ dilute stage 2  structure. The experimental phase diagram between compositions $x\approx0.2$ and 0.5 
appears to be quite complicated. We attribute this to the effects of disorder entropy in the dilute stage 2 structures, to be 
discussed in the next subsection.

Experimentally the phase diagram of Li-graphite is often characterized by measuring the potential difference between a 
Li$_x$C$_6$ electrode and a Li metal electrode,\cite{PhysRevB.44.9170,JES140.2490}
\begin{equation}
V(x) = \frac{1}{e}\left[ \mu_\mathrm{metal} - \mu(x) \right], \label{gibbs1}
\end{equation}
with $\mu_\mathrm{metal}$ the chemical potential of Li metal. The chemical potential of Li in 
Li$_x$C$_6$ $\mu(x)=\partial G(x)/\partial x$ is the derivative of the curve in Fig.~\ref{Li-interc}(b). Because of the convex 
shape of this curve, $V$ is a monotonically decreasing function of $x$. In particular, if at any concentration $x$ two 
stable phases $x_1 < x_2$ are in equilibrium, then the chemical potential is constant in this concentration range, and is given 
by the slope of the corresponding straight line segments in Fig.~\ref{Li-interc}(b)
\begin{equation}
\mu(x) = \frac{G(x_2)-G(x_1)}{x_2-x_1}; \; x_1 \leq x < x_2, \label{gibbs2}
\end{equation}
The potential $V(x)$ as a function of concentration is then a staircase, where each plateau characterizes a mixture of the stable 
compositions $x_1$ and $x_2$. The calculated potential for the Li-graphite system at $T=0$ is plotted in Fig.~\ref{Li-interc}(c). 
Note that the first plateau after starting the intercalation should correspond to minus the intercalation energy $-E_\mathrm{int}$ 
of the first stable phase, which is LiC$_{32}$ ($x=3/16$), cf. Eq.~(\ref{eq:Gx})-(\ref{gibbs2}). The calculated sequence of 
voltage plateaus then follows the sequence of mixtures of stable phases discussed above. 

Compared to the voltages measured in experiment,\cite{PhysRevB.44.9170,JES140.2490,JES151.A422,JES160.A3198,J.Phys.Chem.Solids.57-775,JPCC112.3982} 
the calculated voltages are somewhat too high, e.g., by $\sim 50$~mV at $x=1$. The difference $\Delta V \approx -120$~mV 
between  the $x<3/16$ and the $x>1/2$ plateaus, however, agrees quite well with experiment, suggesting that the calculated results include 
a constant offset. Again, part of this might be due to an error made in the description of Li metal. The shape 
of the voltage curve for $x<0.5$ is quite different from experiment. We attribute this to the effects of finite temperature, as will be discussed in the next subsection.


\subsubsection{Finite temperature}
\label{subsectionfinitetemperature}

As already mentioned, the configurational entropy is zero for the fully loaded stage~1 (LiC$_6$) and stage~2 (Li$_{0.5}$C$_6$) structures.
The configurational entropy contribution to the intercalation free energy could be important, however, for 
the partially loaded stage~1 and stage~2 compounds in Table~\ref{Li-GIC} and Figure~\ref{Li-interc}(a-c). In this section we assess its effect.
We ignore vibrational contributions to energy and entropy as they are only weakly dependent on composition.

To account for the configurational entropy $S^{\rm config}$, we follow the Bethe-Peierls method of Ref.~\onlinecite{JPCC112.3982} and treat 
the intermediate Li$_x$C$_6$ compositions as alloys of occupied and unoccupied Li lattice sites. Positions above the centers of C$_6$ hexagons count as possible lattice sites for Li atoms, as in the structures of Fig.~\ref{Li-struct}, for example. An effective short range repulsion between Li atoms is introduced by excluding configurations where two Li atoms occupy two adjacent, i.e., edge sharing, hexagons, because that is energetically highly unfavorable.\cite{notenn} No longer range interactions between Li atoms are assumed, which means that we probably slightly overestimate the configurational entropy contribution to the free energy. The Bethe-Peierls model provides an exact statistical treatment of Li atoms occupying 7~sites in a hexagonal lattice (the central site and its first ring of neighbors). A mean field treatment accounts for the interactions with the rest of the lattice.

Subtracting $T S^{\rm config}$ from $E_{\rm int}$ and $x T S^{\rm config}$ from $G_0$, Eq.~(\ref{eq:Gx}), we obtain the plots shown in Fig.~\ref{Li-interc}(d-f), calculated for
room temperature ($T=300$~K). Comparing Figs.~\ref{Li-interc}(a) and (d) one observes that the configurational entropy contribution substantially lowers the intercalation (free) energy for some compositions. This also has a marked effect on the Gibbs free energies, shown in Fig.~\ref{Li-interc}(e), where several intermediate compositions besides the $T=0$ structures for $x=3/16, 1/2$ and $1$, are stabilized at $T=300$~K. Constructing the convex curve connecting the free energy minima
we find stable compositions at $x = 3/20, 3/16, 1/4, 3/8, 1/2, 21/24$ and~1.

The calculated potential $V(x)$ at T = 300~K is plotted in Fig.~\ref{Li-interc}(f). Comparing to the situation at $T=0$, Fig.~\ref{Li-interc}(c), we observe that the steps at $x=3/16$ and $x=1/2$ remain prominent. The difference between the plateaus at $x<3/16$ and $x>1/2$ increases somewhat, from $\Delta V \approx -120$~mV ($T=0$) to $-160$~mV ($T=300$~K). The main difference lies in the shape of the curve for intermediate compositions $3/16<x<1/2$, where entropy effects at finite temperature lead to a decrease of the potential step at $x=1/2$ and a concomitant increase of the step at $x=3/16$. Intermediate compositions are also stabilized, but only lead to small potential steps, indicating that the Gibbs free energy $G(x)$ of the (disordered) dilute stage~2 compound is nearly linear in $x$ in this range. The potential rises again at $x\leq3/20$, but for smaller $x$ the calculations become increasingly more difficult.

The voltage curve shown in Fig.~\ref{Li-interc}(f) is in line with what is found in experiments, where a small potential step is typically observed at $x=0.5$, a larger one at or close to $x=0.2$, and further increases of the potential for smaller $x$.\cite{PhysRevB.44.9170,JES140.2490,JES151.A422,JES160.A3198,J.Phys.Chem.Solids.57-775,JPCC112.3982} Evidently including entropy effects leads to a decent description of the voltage curve. It also implies that the curve for $x\leq 0.5$ can be interpreted on the basis of stage 2 compounds only, and that there is no need to invoke stage $N>2$ compounds.

Entropy effects were also studied in Ref.~\onlinecite{JPCC112.3982}, for stage 1 compounds with compositions in the range $0.5 \le x \le 1$, where vdW contributions are likely to be less important. That study employed LDA and GGA functionals without vdW corrections, and found a stabilization at 300~K of the two compositions $x \approx 0.55$ and $x \approx 0.88$.  With the vdW functional we find the composition $x = 0.875$ stabilized at $T=300$~K, but we have not considered structures with compositions near $0.55$. In view of the different functionals used, we consider this good agreement. The calculated potential step at $x = 0.875$ is small, see Fig.~\ref{Li-interc}(f), and it hardly changes the potential curve, as compared to the zero temperature curve, see Fig.~\ref{Li-interc}(c).


\section{Summary and conclusions}
\label{discussion_and_conclusions}

Li/graphite is the archetypical intercalation system. As a material it is of utmost importance for 
applications in rechargeable Li-ion batteries. It shows a remarkable palette of structures and phases as a function of the 
Li concentration Li$_x$C$_6$, $0<x\leq1$. Accurately modeling layered intercalation compounds from first principles has hitherto been very difficult,
as their structure is often determined by a fine balance between van der Waals (vdW) interactions and chemical 
or Madelung interactions, and standard first-principles techniques lack a good description of vdW interactions.

Using recently proposed vdW density functionals we study the structures and the energetics of bulk graphite and Li-graphite 
intercalation compounds. Different versions of  vdW functionals are benchmarked on bulk graphite, where they give a good 
description of the bonding and the structural properties. Selecting the functional that yields the most accurate 
structure (optB88-vdW) one also finds an accurate description of the graphite phonon band structure and the elastic constants 
from first principles.

Intercalation of Li in graphite leads to stable systems with calculated intercalation energies of $-0.2$ to $-0.3$~eV/Li 
atom (referred to bulk graphite and Li metal). The calculations give negative intercalation entropies of $-0.06$ 
to $-0.08$ meV/K/Li atom at room temperature resulting from the phonon contributions, demonstrating that the motion of Li 
atoms in the intercalated compound is more constrained than in the bulk Li metal. 

The fully loaded stage 1 and stage 2 compounds LiC$_6$ and Li$_{1/2}$C$_6$ are thermodynamically stable, corresponding to 
two-dimensional $\sqrt3\times\sqrt3$ lattices of Li atoms intercalated between each pair of graphene planes, or every 
other pair, respectively. Stage $N>2$ compounds, consisting of a $\sqrt3\times\sqrt3$ lattice of Li atoms intercalated between two graphene planes alternating with $N-1$ empty 
layers, are predicted to be unstable. Instead, upon decreasing the Li concentration it is more advantageous to decrease the packing of Li atoms in the stage 2 compound. The compound Li$_{3/16}$C$_6$ is particularly stable; it corresponds to a $\sqrt7\times\sqrt7$ in-plane packing of Li atoms.

Apart from a short-range repulsion the effective in-plane interaction between Li atoms in stage 2 compounds is relatively weak. At elevated temperatures dilute stage 2 compounds Li$_x$C$_6$, $x < 0.5$ are therefore easily disordered. Even at room temperature the relative stability of the Li$_{3/16}$C$_6$ and Li$_{1/2}$C$_6$ structures can still be recognized, however. The voltage profile extracted from the calculations is in reasonable agreement with experiments, which demonstrates the improvements of first-principles techniques in calculating the properties of intercalation compounds.

\section*{ACKNOWLEDGMENTS}
We thank Dr.~J.-S. Filhol for making available his notes on configurational entropy calculation.
The work of EH is part of the Sustainable Hydrogen program of Advanced Chemical Technologies for
Sustainability (ACTS), project no.~053.61.019. The work of GAW is part of the Foundation for Fundamental Research on Matter (FOM) with
financial support from the Netherlands Organisation for Scientific Research (NWO).

\bibliography{Li_gr_resub}

\begin{thebibliography}{82}
\bibitem[{not()}]{note.volume}
\bibinfo{note}{Both cell shape and positions were relaxed for a set of fixed
  volumes. The final equilibrium volume and the ground state energy were
  obtained by a fit to an equation of state.}
\bibitem[{Not()}]{Note.cells}
\bibinfo{note}{Supercells were used: For $\Gamma$-A the cell was doubled along
  $c$. For $\Gamma$-M, we used a $4 \times 4 \times 1$ cell, i.e.\ 4 times
  repeated along $a$ and along $b$. The phonons for the line $\Gamma$-K-M were
  obtained from an orthorhombic supercell with sides $8a \times \sqrt{3} a
  \times c$.}
\bibitem[{fus()}]{fusionconvert}
\bibinfo{note}{The experimental enthalpy of fusion for Li at its melting point
  (453.7~K) is 3~kJ/mol.}
\bibitem[{not({\natexlab{a}})}]{note.LDA}
\bibinfo{note}{Calculations using the LDA give intercalation energies for the
  stage 1 and stage 2 compounds LiC$_6$ and LiC$_{12}$ of $-0.446$ and
  $-0.501$~eV/Li, respectively, implying that LDA considerably overestimates
  the Li-graphite interaction.}
\bibitem[{not({\natexlab{b}})}]{notenn}
\bibinfo{note}{For the energy to move a Li from the next-nearest neighbour to
  the nearest neighbour position of another Li in the bilayer graphene we find
  0.26~eV using optB88-vdW. See also Ref.~\protect \onlinecite{JPCC112.3982}.}
\end{thebibliography}

\end{document}